\documentclass[12pt,fleqn]{article}
\usepackage{amssymb,graphicx,epsfig}
\usepackage{lscape,graphics,amsmath,geometry}
\usepackage{latexsym}
\usepackage{cite}

\newcommand{\lsim}{\stackrel{<}{_\sim}}
\newcommand{\gsim}{\stackrel{>}{_\sim}}

\newcommand{\met}{\mbox{\ensuremath{\slash\kern-.7emE_{T}}}}
\newcommand{\mpt}{\mbox{\ensuremath{\slash\kern-.7emp_{T}}}}
\newcommand{\me}{\mbox{\ensuremath{\slash\kern-.7emE}}}
\newcommand{\missp}{\mbox{\ensuremath{\slash\kern-.7emp}}}

\begin{document}

\pagestyle{empty}

\vspace*{2.0cm}

\noindent
DESY 07-135 \\
September 2007 
\vspace*{0.6cm}

\begin{center}

{\Large\sc {\bf Prospects of a Search for a New Massless Neutral Gauge
Boson at the ILC}}
\vspace*{1.2cm}

{\sc E. Boos}$^{1}$, {\sc V. Bunichev}$^{1}$ and {\sc  H.J. Schreiber}$^{2}$
\begin{small}
\vspace*{1.0cm}

$^1$  Skobeltsyn Institute of Nuclear Physics, MSU, 119992 Moscow, Russia \\
\vspace{3mm}
$^{2}$ DESY, Deutsches Elektronen-Synchrotron, D-15738  Zeuthen, Germany

\end{small}
\end{center}
\vspace*{0.7cm}

\pagestyle{plain}

\pagenumbering{arabic}


\vspace{1.0cm}
\begin{center}
\section*{Abstract}
\end{center}
\noindent
Prospects to search for a new massless neutral gauge boson, the paraphoton,
in $e^+e^-$ collisions at center-of-mass energies of 0.5 and 1 TeV are studied. 
The paraphoton naturally appears in models with abelian kinetic mixing.
A possible realistic
model independent lowest order effective Lagrangian contains magnetic
interactions of the paraphoton with the Standard Model fermion fields.
These interactions are proportional to the fermion mass and grow with
energy, with however very weak paraphoton couplings to ordinary matter.
At the ILC, a potentially interesting process to search for the
paraphoton is its radiation off top quarks, so that
the event topology to be searched for is a pair of acoplanar top quark jets 
with missing energy. 
By combining many discriminating features of signal and background events
efficient paraphoton event selection was achieved allowing to set
limits for the top-paraphoton coupling. Arguments in favor 
of the missing energy as the paraphoton with spin 1 are discussed.

\newpage


\section{Introduction}
Modern elementary particle field theories are based on principe of the
gauge invariance. It means that the Lagrangian of the theory should be
invariant with respect to group transformation of the local symmetry which leads
to a corresponding number of massless vector gauge boson fields.
In the Standard Model (SM), based on the $U_Y(1) \times SU_L(2) \times SU_C(3)$
gauge symmetry group, 12 gauge vector bosons exist.
Three of them, the electroweak bosons $W^{\pm}$ and $Z^0$, get masses due to the Higgs
mechanism of spontaneous symmetry breaking. The eight massless
strongly interacting gauge bosons, the gluons, are confined in hadrons
and only one directly observed
massless neutral vector boson, the well known photon, exists  within the SM.

Although the Standard Model does not require any additional
gauge fields it is possible to introduce gauge invariant
operators in the Lagrangian which involve new gauge
fields not forbidden by basic principe of gauge invariance.
An example is given in \cite{Holdom:1985ag} by the abelian kinetic mixing
of the SM $U_Y(1)$ field with a new $U_P(1)$ field in a gauge invariant manner. 
The mixing term of the two $U(1)$ fields 
can be diagonalized and canonically normalized by an $SL(2,R)$
transformation in a way that one linear combination
of the fields corresponds to the ordinary photon which
couples in the usual manner to all electrically charged particles within the SM.
The other linear combination appears as a massless spin-1 neutral particle,
referred to as the 'paraphoton' in \cite{paraphoton}
and denoted by $\gamma^{\prime}$ in this paper.
This mechanism also provides an elegant way
of introducing millicharged particles\footnote{ Millicharged particles
are related to fields charged under the $U_P(1)$ group.
Interaction of these particles with SM fields should be very small and 
proportional to the kinetic mixing parameter.}
into the theory \cite{millicharge}. 
The paraphoton couples directly to millicharged fermions and only 
indirectly to the SM fields
via higher mass-dimension operators.

In this study we follow an approach proposed in \cite{Dobrescu:2004}
where  the effective Lagrangian of the interaction of the paraphoton
with the SM fermion fields was proposed by considering
higher dimensional operators.
A possible lowest order Lagrangian 
which preserves both the new $U_P(1)$  and the SM gauge symmetries
with the SM fermion cirality structure has the following form:
\begin{eqnarray}
\frac{1}{M^2} P_{\mu\nu} \left(\bar{q}_L \sigma^{\mu\nu} C_u \tilde{H} u_R
+ \bar{q}_L \sigma^{\mu\nu} C_d H d_R + \bar{l}_L \sigma^{\mu\nu} C_e H e_R + h.c\right),
\end{eqnarray}
where $q_L,l_L$ are the quark and lepton doublets, $u_R,d_R$ the up and
down-type $SU(2)$ singlet quarks, $e_R$ the electrically-charged
$SU(2)$-singlet leptons, and $H$ is the Higgs doublet. An index
labeling the three fermions generations is implicit here.
The $3 \times 3$ matrices in flavor space, $C_u,C_d,C_e$, have dimensionless complex
elements, and $M$ is the mass scale where the operators are generated.

One can see that the interactions of the paraphoton with Standard
Model fermions are suppressed by two powers of the mass scale $M$, but are
directly proportional to the fermion mass $m_{f}$ 
and the dimensionless coupling strength
parameter $C_f$, with $f=u,d,e$. The coefficients $C_f$ are unknown,
but various phenomenological constraints exist. 
In particular, limits on these parameters for light fermions were,
for example, deduced from paraphoton annihilation to muon pairs,
$\gamma^{\prime} \gamma \to \mu^+ \mu^-$,
or the Compton-like process, $\gamma^{\prime} \mu^{\pm} \to \gamma \mu^{\pm}$,
assuming the $\gamma^{\prime}$ interaction rate equals the expansion rate
of the universe at freeze out.
Together with  successful predictions of primordial nucleon-synthesis the
$\mu^-\mu^+\gamma^{\prime}$ coupling parameter is bounded 
to $M/\sqrt{c_{\mu}} \gsim 1.5$ TeV,
where $c_{\mu}$ is related to $C_{\mu}$ via $c_f=C_f v_h/(\sqrt{2}m_f)$,
with $v_h$ the vacuum expectation value of the Higgs field.
Or, star cooling by $\gamma^{\prime}$ emission constraints the electron-paraphoton
interaction since the associated energy loss
is proportional to the square of $4 c_e {m_e}^2/M^2$. 
The limit on $\gamma^{\prime}$ emission through Bremsstrahlung,
such as $e^- + {}^4He \to e^- + {}^4He + \gamma^{\prime}$, from the core of
red giant stars \cite{Pospelov:2000bw} requires
$M/\sqrt{c_e} \gsim 3.2$ TeV, while Compton-like scattering,
$\gamma e^- \to \gamma^{\prime} e^-$, in horizontal-branch stars 
sets a somewhat weaker limit of
$M/\sqrt{c_e} \gsim 1.8$ TeV. A constraint on the $\gamma^{\prime}$-coupling 
to nucleons of $M/\sqrt{c_N} \gsim 7$ TeV has been estimated from
the neutrino signal of the supernova 1987A assuming the supernova was
cooled predominantly by neutrinos. 
More details of possible lower limits on $\gamma^{\prime}$ interactions
with fermions are discussed in \cite{Dobrescu:2004}.

An intriguing aspect of the presence of an additional gauge field
like the paraphoton is the possible existence
of fields charged under the $U_P(1)$ group. Simple renormalizable models
generate operators, see eq.(1), which are associated with new heavy states.
The lightest particle of this type with negligible electrical charge 
is stable and could be a viable dark matter candidate.

To summarize, a massless neutral gauge boson other than the SM photon
may exist. It interacts with ordinary matter via higher-order operators.
The rather weak bounds on the mass scale M
makes it worthwhile to search for this new photon-like
state in future collider experiments.
From the Lagrangian, eq.(1), follows that due to the proportionality of the
$\gamma^{\prime}$  couplings to the fermion mass,
$\gamma^{\prime}$ interaction with SM particles is strongest with 
particles of the third generation, especially with the top quark,
and small or negligible with light fermions.
Therefore, we expect that the most interesting process to search
for the paraphoton will be $\gamma^{\prime}$ radiation
off the top quark. Since so far no constraint on $c_t$ exists,
access to $M/\sqrt{c_t}$ seems possible or corresponding
limits might be set for the first time.

It seems a priori very difficult to perform $\gamma^{\prime}$ searches at hadron
colliders because of copious $ t\bar{t}$ + multi-jet background production.
The next generation $e^+e^-$ linear collider (ILC) is ideally suited 
to evaluate prospects of a search for the paraphoton via the channel
\begin{eqnarray}
e^+e^- \to t~\bar{t}~\gamma^{\prime}~.
\end{eqnarray}
The search strategy relies on the property of the $\gamma^{\prime}$
to interact weakly with ordinary matter and its favored
emission from top quarks.
Hence, the signal signature consists
of a pair of acoplanar top quark jets with missing transverse energy,~\met,
carried away by the paraphoton. 
The rate of such events if noticed
should clearly exceed the expected SM background.

Simulations of $t\bar{t}\gamma^{\prime}$ signal events
with a 'reasonable' value of the
coupling parameter $M/\sqrt{c_t}$ and SM background reactions were performed at
center-of-mass energies $\sqrt{s} = 0.5 $ and 1.0 TeV 
and an integrated luminosity of 0.5, respectively, 1 ab$^{-1}$.
These assumptions are in accord with the present design for the ILC,
initially producing collisions at 0.5 TeV and in a second stage at 1 TeV \cite{RDR}.

The paper is organized as follows. In section 2 basic properties of
signal events are studied at the parton level in order to extract informations
which might be helpful to discriminate signal from background events.
Section 3 describes the search strategy for the $\gamma^{\prime}$ based on
hadronic $W$ decays, $W \to q \bar{q}$, to avoid
complications from leptonic $W$ decays with neutrinos in the final state,
also carrying away transverse energy. The analysis is performed
based on full simulation including ILC detector response.
Section 4 discusses accessible limits on $M/\sqrt{c_t}$ from excess
of signal events over the SM background expectations
and arguments in favor of the nature of the paraphoton are presented.
Conclusions are summarized in section 5.

%
%
\section{The signal reaction $e^+e^- \to t~\bar{t}~\gamma^{\prime}$}

The characteristics of the signal reaction $e^+e^- \to t~\bar{t}~\gamma^{\prime}$
were computed and corresponding partonic events were generated by means of
the program package CompHEP \cite{Pukhov:1999gg}.
The Feynman rules for the fermion-fermion-$\gamma^{\prime}$  vertices
following from the effective Lagrangian (1)
\begin{eqnarray}
\frac{c_f}{M^2}\cdot m_f \cdot p_{\nu}^{\gamma^{\prime}}
\big(\gamma^\nu \gamma^\mu -\gamma^\mu \gamma^\nu \big) 
\end{eqnarray}
have been implemented into CompHEP allowing variations of 
the free coupling parameter $M/\sqrt{c_t}$.
An interface with PYTHIA 6.202 \cite{Sjostrand:2003wg} 
simulates initial and final state radiation
and jet hadronization, needed at a later stage of our study. Also,
beamstrahlung effects \cite{Aguilar-Saavedra:2001rg} are taken into account.

Table 1 shows 
the number of signal events expected at $\sqrt{s} = 0.5$ and 1 TeV
as a function of $M/\sqrt{c_t}$ for an accumulated luminosity of 0.5,
respectively, 1 ab$^{-1}$.
The event rates become rapidly smaller
with increasing $M/\sqrt{c_t}$, so that in particular 
for large $M/\sqrt{c_t}$ values $\gamma^{\prime}$ detection is challenging.
Simulations were only performed for $M/\sqrt{c_t}$ = 0.2 TeV
enabling sufficient $\gamma^{\prime}$ events at both energies.
It is worth to mention that below $M/\sqrt{c_t} \simeq$ 0.1 TeV 
the kinetic mixing parameter of the $U_Y(1)$ and $U_P(1)$
becomes relatively large and leads to millicharged particles
with electric charges larger than $10^{-4}$. Such  particles should have been detected
in various SM reactions, contrary to experimental findings.
One should also point out that the effective interaction Lagrangian approach
cannot be applied for $M/\sqrt{c_t} \lsim $ 0.1 TeV 
since the effective coupling parameter $\frac{c_t m_{top} E} {M^2}$
gets to large for typical collision energies $E$.

\begin{table}
\begin{center}
\begin{tabular}{|c|c|c|} \hline
$M/\sqrt{c_t}~~[TeV]$ & $\sqrt{s}=0.5~TeV$ & $\sqrt{s}=1~TeV$\\ 
\hline
$0.2$ & $5 700$ & $42 500$ \\
\hline
$0.3$ & $1 100$ & $8 500$ \\
\hline
$0.5$ & $40$ & $1 100$ \\
\hline
$1$ & $10$ & $70$ \\
\hline
\end{tabular}
\end{center}
\caption{ $t\bar{t}\gamma^{\prime}$ event rates for several values
  of $M/\sqrt{c_t}$
  at $\sqrt{s} = 0.5$ and 1 TeV and
  an integrated luminosity of 0.5, respectively, 1 ab$^{-1}$.}
\end{table}

\begin{figure}[ht]
\begin{center}
\includegraphics[width=74mm,height=50mm]{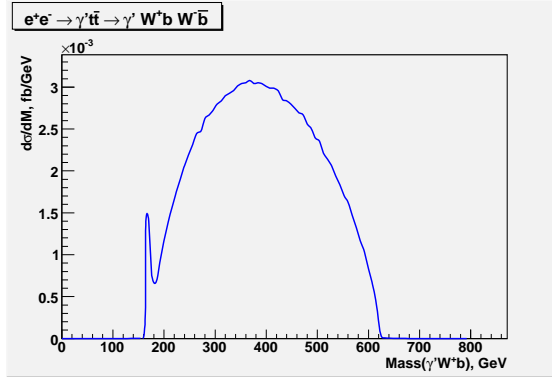}
\caption{Invariant mass of the $\gamma^{\prime} W b$ system.}
\label{inv_mass}
\end{center}
\end{figure}

\begin{figure}[ht]
\centering
\includegraphics[width=74mm,height=50mm]{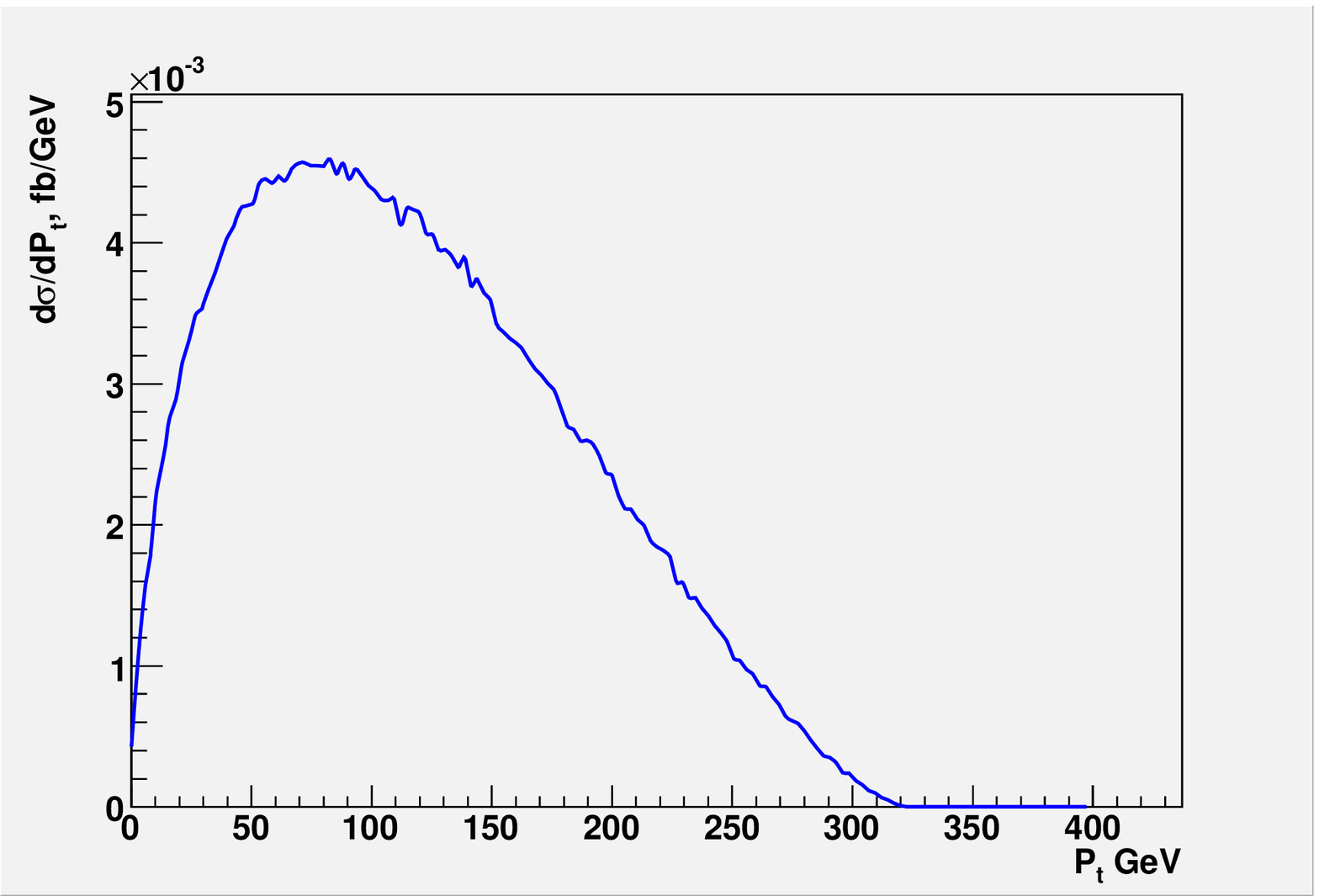}
\includegraphics[width=74mm,height=50mm]{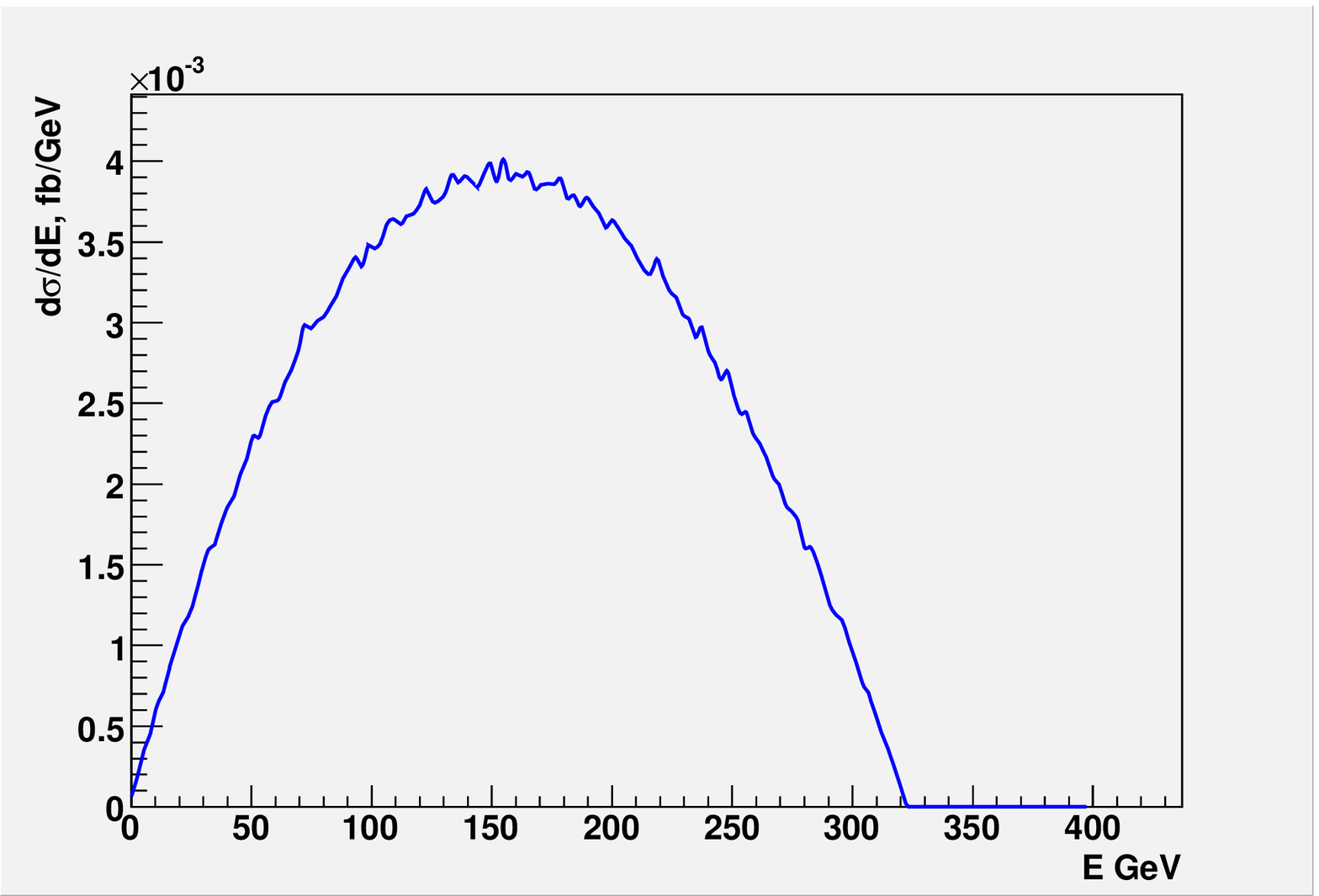}
\caption{$\gamma^{\prime}$ transverse momentum (left) and energy (right) 
  distributions at $\sqrt{s} = 1$ TeV.}
\label{e_miss}
\end{figure}

In order to establish a search strategy for the paraphoton in $t\bar{t}$ events
it is advantageous to know whether an off-shell or on-shell top quark radiates
the $\gamma^{\prime}$. Fig.~\ref{inv_mass}
shows the invariant mass of the $\gamma^{\prime} W b$ system of that top
which radiates the paraphoton. 
Besides of a tiny fraction of events with a $\gamma^{\prime} W b$ invariant mass
at the top mass, most of the events have a $\gamma^{\prime} W b$ mass
larger than $m_{top}$.
These events have the corresponding $Wb$ invariant mass close to the top mass.
Therefore, the paraphoton is radiated off a top being off-shell in most 
cases, and $\gamma^{\prime}$ search strategies should rely on an
on-shell top with $t \to W b$ decay in association with the $\gamma^{\prime}$.
Since the signature of the $\gamma^{\prime}$
relies on negligible interaction strength with ordinary matter,
its transverse momentum and energy behavior is essential for its finding.
Fig.~\ref{e_miss} shows both variables at 1 TeV.
Note that substantial transverse momentum, respectively,
energy is carried away by the $\gamma^{\prime}$, so that
large missing transverse momentum, ~\mpt,
and/or missing energy, ~\me, will tag signal events.

%
%
\section{Signal event selection}

After event generation using CompHEP, PYTHIA
and the CompHEP-PYTHIA interface with the Les Houches Accord implemented 
\cite{Boos:2001cv},
an approximate response of an ILC detector was simulated
with the package SIMDET$_{-}$v4 \cite{Pohl:2002vk}
resulting to 'measured' tracks and energy clusters in the calorimeters.
Including a simple particle flow algorithm, the output of SIMDET
denoted as 'energy flow objects' was subject to our search studies.

\begin{figure}[ht]
\centering
\includegraphics[width=74mm,height=50mm]{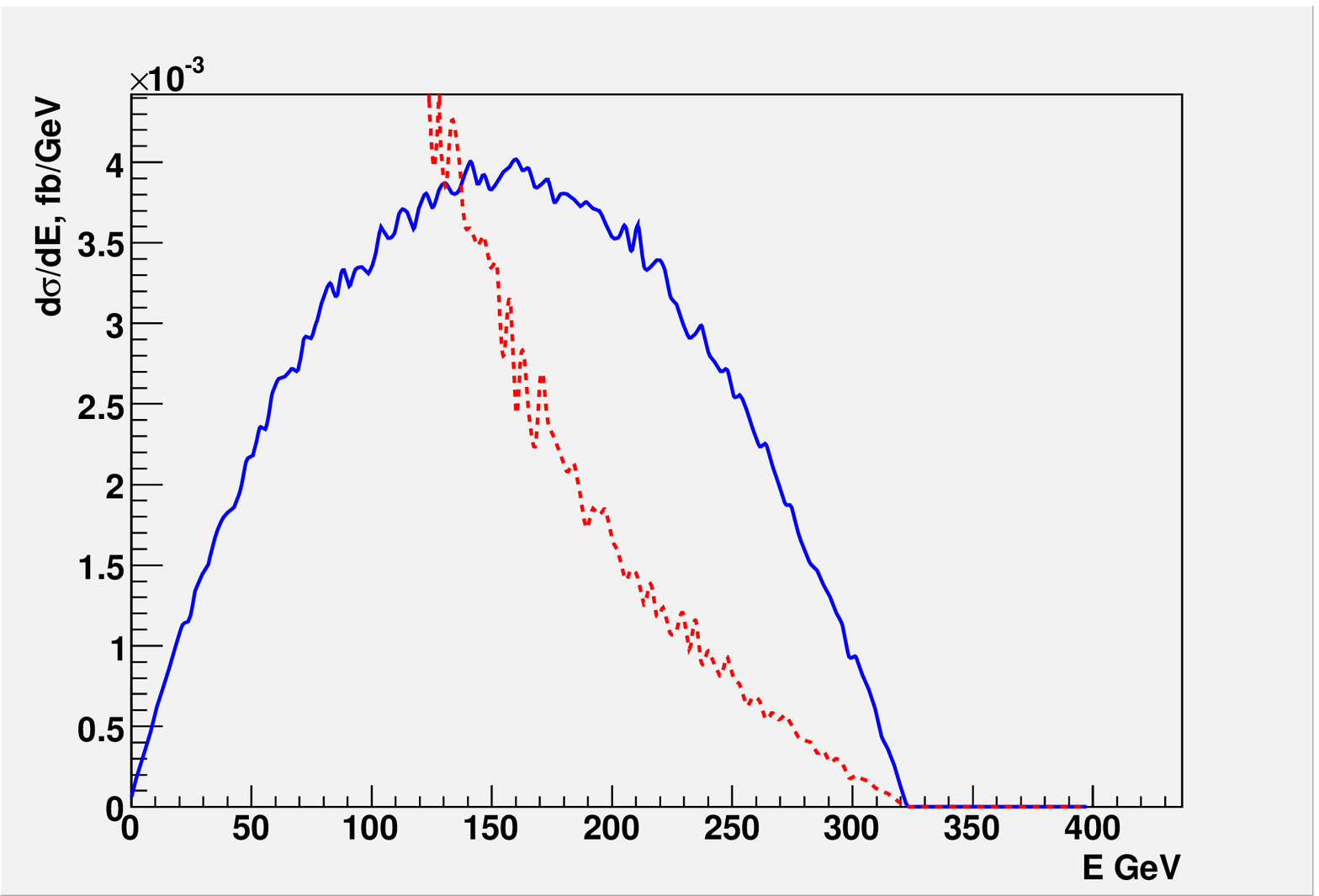}
\includegraphics[width=74mm,height=50mm]{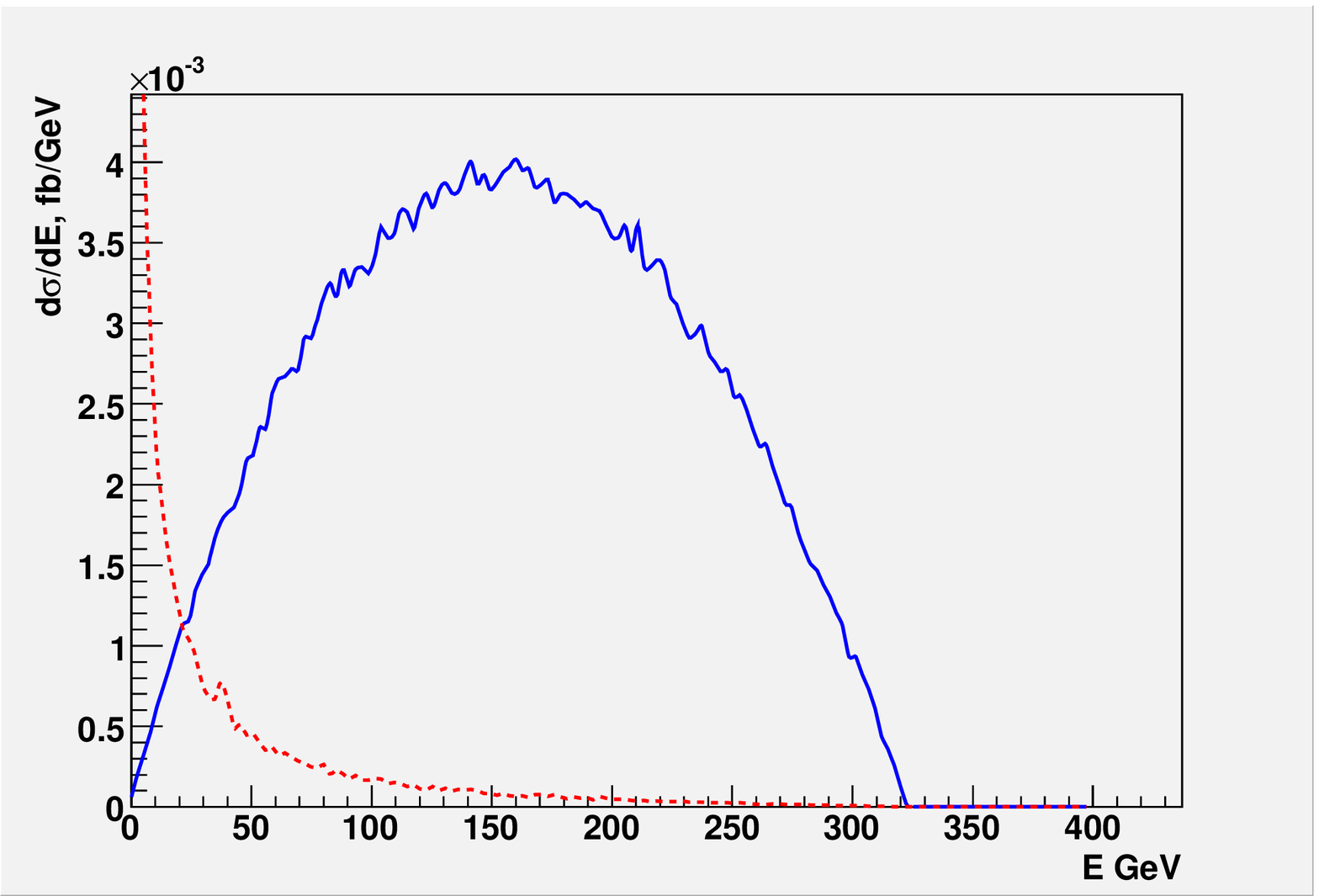}
\caption{Energy spectrum of the $\gamma$ (dotted line) from $t\bar{t}+(\gamma)$
  background and the $\gamma^{\prime}$ (solid line) from $t\bar{t}\gamma^{\prime}$ signal
  events. The left side shows the spectra without any cut, the right one with the cut
  $cos(\theta_{\gamma}) > 0.95$.}
\label{photon}
\end{figure}

Basic properties of the signal process
as discussed in the previous section may suggest
that a reasonable separation of $t\bar{t}\gamma^{\prime}$ events 
from large SM background should be possible.
However, there are a number of SM background sources which have similar or
identical final state signatures, i.e. a signature of $t\bar{t}$ + \met 
~with acoplanar top quark jets.
The most important background consists of ~$t\bar{t}+(\gamma)$ events,
where the $\gamma$ from initial state radiation (ISR) is very often not detected.
The number of events expected for both energies are given in Table 2.
They exceed substantially
the number of signal events for interesting $M/\sqrt{c_t}$ values.
Fig.~\ref{photon} shows the photon energy for such events without cuts
and with the demand $cos(\theta_{\gamma})>0.95$
in comparison to the $\gamma^{\prime}$ distribution from signal events.
\begin{table}
\begin{center}
\begin{tabular}{|c|c|c|} \hline
$background$ & $\sqrt{s}=0.5~TeV$ & $\sqrt{s}=1~TeV$\\ 
\hline
$t\bar{t}(\gamma)$ & $276 675$ & $200 310$ \\
\hline
$t\bar{t}\nu\bar{\nu}$ & $75$  & $930$ \\
\hline
\end{tabular}
\end{center}
\caption{Background events at $\sqrt{s} = 0.5$ and 1 TeV for an
integrated luminosity of 0.5, respectively, 1 ab$^{-1}$.}
\end{table}
\hspace{-4mm}As can be seen, the paraphoton energy spectrum is very different
from the background photon spectrum. The $\gamma^{\prime}$ energy
rises due to the momentum proportionality of its coupling with the top quark.
In contrast, the ISR photon spectrum peaks at zero, which makes
the photon energy a potentially sensitive dicsriminating variable.
Obviously, a veto on registered photons with small polar angle would 
discard a substantial fraction of this background. However, in practice
photons with very low $\theta_{\gamma}$ are not accessible and
copious small polar angle $\pi^0$ and $\gamma$ production from quark fragmentation
occurs, and a veto on final state photons would also eliminate  
signal events. Hence, such a cut will not be applied in this study.

The next significant background to consider is
the channel $e^+e^- \to t\bar{t}+\nu\bar{\nu}$,
with a signature similar to that of the
signal due to escaping neutrinos in the final state,
with $\nu\bar{\nu}$ pairs coming mostly from Z decays.
The corresponding event numbers also given in Table 2 are
comparable to the signal event rates
for not too small $M/\sqrt{c_t}$ values, but orders of magnitude smaller
than the dominant $t\bar{t}(\gamma)$ background.
An invariant mass cut of e.g. $M_{\nu\bar{\nu}} <$ 80 GeV,
i.e. a cut on the event missing mass,
removes most of the $t\bar{t}+\nu\bar{\nu}$ events.

Additional SM background with a topology of a pair of acoplanar top quark jets
and large ~\met ~is not needed to be addressed. 
It will be difficult for any SM event to mimic the topology of interest,
so that the only source of potential background
to consider consists of $e^+e^- \to t~\bar{t}~(\gamma)$ events.
It is important to have some good understanding of this reaction in order to establish
an excess of paraphoton candidate events over the background.

In a first attempt, a conventional method was applied
by using consecutive cuts on kinematical variables based on either
the energy flow objects or, utilizing a jet finder, the 4-momenta
of jets consistent with the
$t\bar{t} \to (Wb)(Wb) \to (q\bar{q})b~(q\bar{q})b$ decay chain.
The variables used may be classified into three categories:
global event kinematics, variables based on jet properties
and variables based on jet correlations. We considered
the missing event energy, ~\me, missing transverse energy, ~\met,
missing momentum, ~\missp, missing transverse momentum, ~\mpt, the event aplanarity,
thrust, missing mass squared, the angle between the top momenta
and the coplanarity angle (the angle between the beam, the $t$ and $\bar{t}$)
as well as, for a given hemisphere,
the largest energy of jets and the largest angle between two jets.
Jets were reconstructed by means of
the routine PUCLUS from PYTHIA which relies on a cluster analysis method
using particle momenta. The 'jet-resolution-power' was adjusted to provide
7- and 8-jet event rates in accord with expectations from
gluon radiation. The method of consecutive cuts, however, was found to be inefficient
to select signal from background because of the failure of distinct properties
between signal and background events.

\begin{figure}[ht]
\centering
\vspace{-45mm}
\includegraphics[width=150mm,height=120mm]{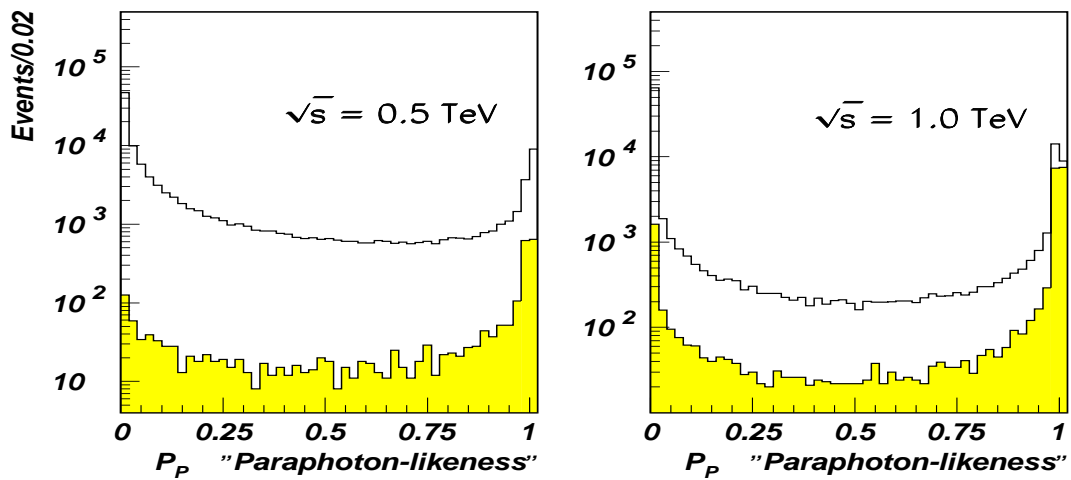}
\caption{Distributions of the discriminant variable~$P_P$ for $t\bar{t}\gamma^{\prime}$ 
signal events (shaded) and the sum of signal and background events 
   at $\sqrt{s} = 0.5$ (left) and 1 TeV (right).}
\label{paraphoton_prob_all}
\end{figure}

In cases of large background, small signal event rates and of variables
with only small discrimination power one needs to pursue more sophisticated strategies
to extract the signal. 
Out of several powerful multivariate selection methods we used the following.
Kinematical variables as discussed above
were combined into a global discriminant variable~$P_P$, designed to give
a measure of the 'Paraphoton-likeness' of any particular event.
This quantity was constructed from the variables after normalization
based on large statistics samples of simulated signal and background events.
For each event and variable $i$, signal and background probabilities
($P_{S}^i$, respectively, $P_{B}^i$) were then calculated,
and by multiplication of all signal probabilities 
($~\prod^n_{i=1}\frac{P_{S}^i} {P_{S}^i + P_{B}^i}$,
$i$=1 ... n, with n=18, the number of variables taken into account) the sensitivity
for an event to be a paraphoton candidate was maximized.
The quantity so obtained was constraint to lie in the region [0;1].
Background events are preferentially  distributed at low $P_P$ values
while for signal events $P_P$ is close to unity.
The distributions of~$P_P$ for both center-of-mass energies considered are shown
in Fig.~\ref{paraphoton_prob_all}. Clear accumulations of $\gamma^{\prime}$ candidate events
can be recognized near $P_P$ = 1, with some non-negligible background 
in particular at 0.5 TeV. A cut of $P_P>0.98$ was applied to select 
signal events.
This method resulted in a $\gamma^{\prime}$ selection efficiency of 49\% (76\%) 
at $\sqrt{s}$ = 0.5 (1) TeV, while only 9\% of background events
at both energies survived. The 'Paraphoton-likeness' technique results 
to significantly better signal-to-background ratio
than the method using consecutive cuts. 
Therefore, we rely on the results of this method and demand $P_P>0.98$
as the principal cut in the study.

At $\sqrt{s}$ = 0.5 TeV after application of the likeness cut,
S/$\sqrt{B}$ results to 11.96 for $M/\sqrt{c_t} = 0.2$ TeV,
while S/$\sqrt{B}$ is 162.6 at 1 TeV, i.e. the chance of measuring the signal event rates
as a result of a background fluctuation is $0.5\cdot10^{-12}$ 
and $< 10^{-15}$ at 0.5, respectively, 1 TeV, using Gaussian sampling of uncertainties.
These numbers clearly demonstrate that background fluctuation cannot 
be responsible for the excess. Including a conservative
3\% uncertainty of the $t\bar{t}(\gamma)$ background rate
would not alter the conclusions on the highly significant excess 
of $\gamma^{\prime}$ events.

\begin{figure}[ht]
\vspace{-15mm}
\centering
\includegraphics[width=140mm,height=110mm]{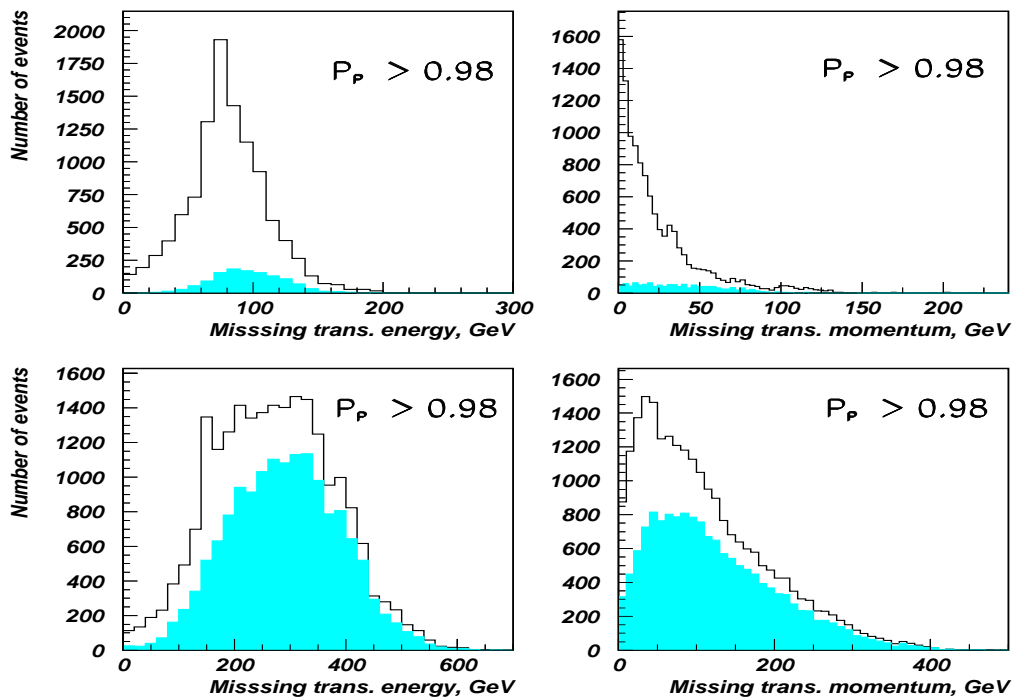}
\caption{\met~and ~\mpt~distributions of $t\bar{t}\gamma^{\prime}$ signal events
   (shaded) and the sum of signal and background events at $\sqrt{s}$ = 0.5 (top)
   and 1 TeV (bottom).}
\label{energy_trans_miss_all}
\end{figure}

Fig.~\ref{energy_trans_miss_all} shows the ~\met ~and ~\mpt ~distributions
at 0.5 and 1 TeV for the signal events (shaded) and the sum of
signal and background events, surviving the cut $P_P > 0.98$.
As apparent from Fig.~\ref{energy_trans_miss_all},
convincing excess of paraphoton events is evident in both distributions
at 1 TeV and the ratio $S/\sqrt{B}$ being in the order of 162
can be further enhanced by demanding, for example,
~\met~$>$ 330 GeV or ~\mpt~$>$ 100 GeV.
In this way, an almost background-free signal event sample can be extracted
for further measurements. The situation is much less
convenient at 0.5 TeV, where reasonable signal event extraction with small background
is difficult to achieve. Improvements are, however, expected for
$M/\sqrt{c_t}$ 'coupling' values less than 0.2 TeV (see Table 1) so that
$\gamma^{\prime}$ physics can also be probed during the first phase of the ILC.

\section{Discussion of the results}

If an excess of signal events over SM background expectations has been established,
limits on $M/\sqrt{c_t}$
accessible for a significance\footnote{ We will quantify the discovery potential
of the $\gamma^{\prime}$ in the usual way of $significance = signal/\sqrt{background}$,
where signal and background imply the number of corresponding events
passing all cuts.} 
of $S/\sqrt{B} = 5$ can be derived. We consider this figure
as sufficient for discovery the paraphoton
characterized by acoplanar top quark jets with large
~\met, carried away by the $\gamma^{\prime}$. The number of surviving
$\gamma^{\prime}$ events for 5$\sigma$ discovery amounts to
508 (450) at 0.5 (1) TeV for an integrated luminosity of 0.5 (1.0) ab$^{-1}$.
These numbers can be converted into a limit for the 'coupling' parameter 
$M/\sqrt{c_t}$ of 0.33 (0.61) TeV, with the 1 TeV value of 0.61 TeV
as the most stringent limit accessible at the ILC.

We will also discuss the signal-to-background ratio, $S/B$,  as it will be important
for attempting to understand the nature of the excess events. For that,
we would like to have a clean sample of events that we understand to be mostly signal.
This is especially important for studying variables in favor of the interpretation
of the missing energy as the paraphoton.
Large numbers of background events
would dilute signal properties and hence complicate correct interpretations.
At $\sqrt{s}$ = 0.5 TeV, the small $S/B$ ratio of 0.11 does not favor such an analysis
despite of small possible improvements of the performance by additional cuts.
At 1 TeV, however, the ratio $S/B$ of 1.79 is sufficiently large
so that background contamination should not be a major worry.
If we require in addition ~\met $>$ 330 GeV, the number of signal
to background events results to 5231/2654 = 1.97.
           
In order to demonstrate the spin-1 nature of the $\gamma^{\prime}$, we follow studies
performed to establish the vector nature of the gluon in 3-jet $e^+e^-$ annihilation events
at PETRA \cite{TASSO_1980,CELLO,PLUTO,MARK_J} and LEP \cite{OPAL,L3,DELPHI} energies,
based on predictions that a spin-$\frac{1}{2}$ quark radiates the spin-1 gluon.
Many observables have been measured, including the Ellis-Karlinger angular distribution
\cite{Ellis}, energy-energy correlations \cite{TASSO_1987}, jet masses \cite{TASSO_THESIS}
as well as the three-jet \cite{TASSO_1985} and multi-jet production cross sections
\cite{TASSO_1988}, all of which were important in establishing
the properties of the gluon, in particular its spin.

\begin{figure}[ht]
\vspace{-25mm}
\centering
\includegraphics[width=100mm,height=100mm]{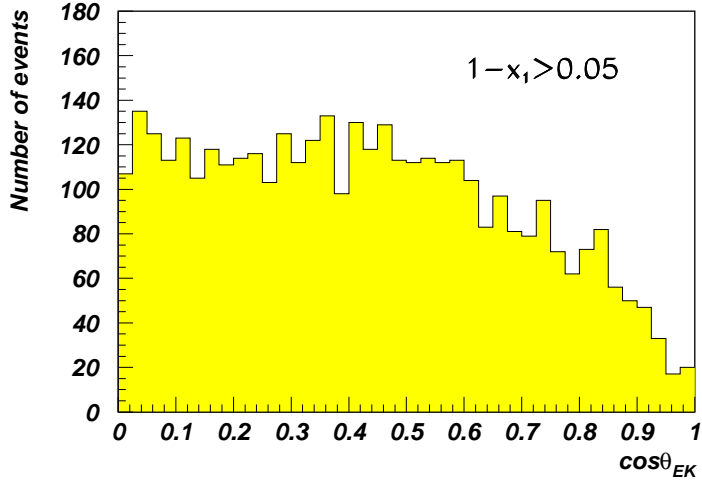}
\vspace{-3mm}
\caption{Cosine of the Ellis-Karliner angle $\theta_{EK}$ for 3-jet events
    selected at $\sqrt{s}$ = 1 TeV with the additional cut $1 - x_1 > 0.05$.}
\label{ellis_karliner}
\end{figure}

For the sake of demonstration, we assign
for each 1 TeV signal event candidate with ~\met $>$ 330 GeV,
the fracional energy variables $x_i = E_i/E_b$ (i=1, 2, 3) to the 
$t,~\bar{t}~$ and $\gamma^{\prime}$, with $E_b$ the nominal incident
beam energy.
 After ordering the $x_i$ such that $x_1 \ge x_2 \ge x_3$ and 
 assuming the top quark mass is small with respect to $E_b$,
 a Lorentz boost is performed  which brings the two less energetic
jets to their c.m. frame where they should emerge back-to-back.
The angle which these jets make with the thrust axis is defined as
the Ellis-Karliner angle $\theta_{EK}$ \cite{Ellis}
\begin{eqnarray}
    cos\theta_{EK} = \frac{x_2 - x_3}{x_1} = \frac{sin\theta_2 - sin\theta_3}{sin\theta_1}~,
\end{eqnarray}
where $\theta_i$ is the angle between the two jets opposite to jet $i$.
Fig.~\ref{ellis_karliner} shows,
after background subtraction and some detector acceptance corrections,
 the cosine of the Ellis-Karliner angle distribution proposed
to dicriminate between the vector and scalar nature of the radiated $\gamma^{\prime}$,
In order to avoid infrared divergences the analysis
is restricted to a region safely away from $x_1=1$, by the cut $1 - x_1 > 0.05$.
Distinction between the vector and scalar particle interpretations is made only on the basis
of the shape of the distribution: a spin-1 particle provides
a flat behaviour near $cos\theta_{EK}$= 0, while a spin-0 object
yields a rising behaviour \cite{TASSO_1980}. Thus,
spin-1 assignment for the paraphoton is highly favored over spin 0.

\begin{figure}[ht]
\centering
\includegraphics[width=160mm,height=120mm]{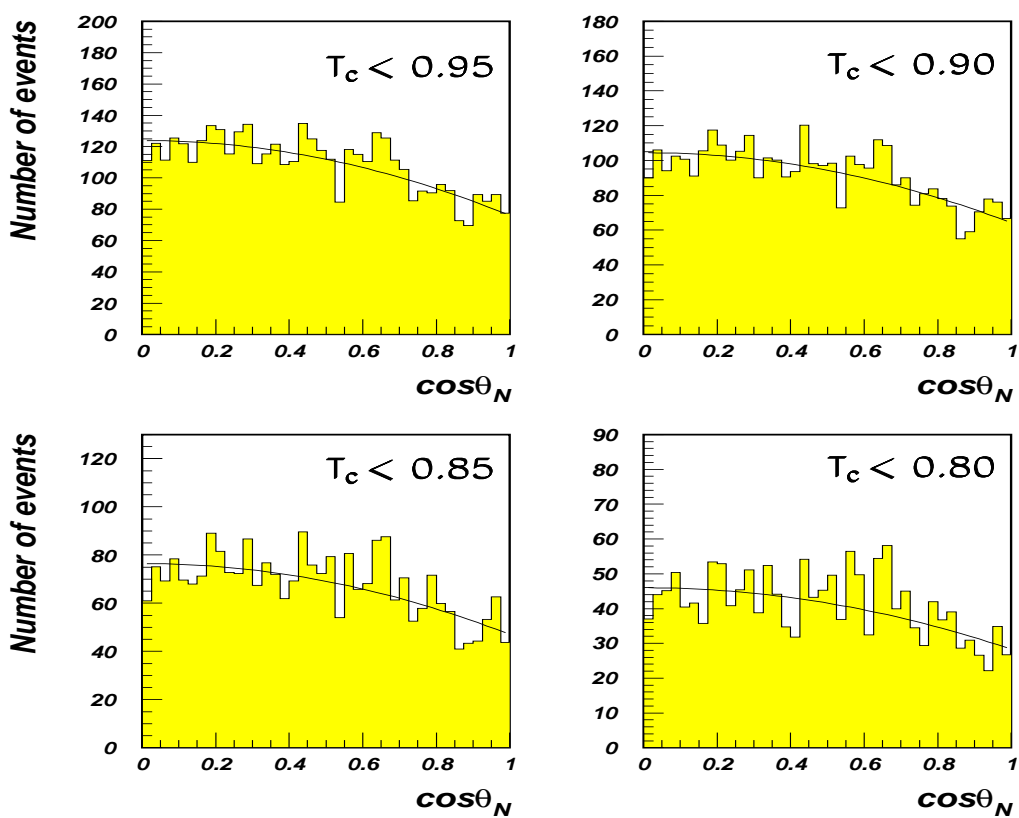}
\vspace{-3mm}
\caption{Polar angle distributions of the normal to the three-jet plane
    for four different thrust cut-off values at $\sqrt{s}$ = 1 TeV.
    The curves represent the results of the fits described in the text.}
\label{alpha_normal}
\end{figure}

Alternatively, after interpreting a signal candidate event as a 3-jet event,
the polar angle distribution of the normal to the three-jet plane, $\theta_N$,
was proposed as a variable to distinguish between the vector
and scalar hypothesis of the emitted particle \cite{Kramer,Koller}.
We discuss shortly this variable which is
defined by the cross-product of the two fastest jets as a function of
a thrust cut-off $T_C$ in order to be able to establish that the parameter $\alpha_N$
extrated from the distributions is (i) independent of the thrust cut-off chosen
and (ii) close to $-\frac{1}{3}$ as predicted for the spin-1 interpretation 
of the $\gamma^{\prime}$.
The cosine distributions of the angle $\theta_N$ are shown in Fig.~\ref{alpha_normal}
for various thrust cut-off values. The distributions, corrected for background
and detector effects, were fitted to the expression
\begin{eqnarray}
  \frac{1}{N} \frac{dN}{dcos\theta_N} = \frac{1}{2(1+\frac{1}{3} \alpha_N)}
     (1+\alpha_N cos^2\theta_N)~,
\end{eqnarray}
predicted for vector particle emission \cite{Koerner} with $\alpha_N = -\frac{1}{3}$.
Good agreement between the data and the theoretical expectation is found.
In particular, as seen from Table 3,
the $\alpha_N$ values are found to be independent of the thrust cut-off
and quite close to $-\frac{1}{3}$, despite of neglecting detailed correction factors
in the analysis.
\begin{table}[ht]
\begin{center}
\begin{tabular}{|c|c|} \hline
$ T_C $ & $ \alpha_N $ \\
\hline
$~~0.95~~$ & $~~-0.386 \pm 0.036~~$ \\
$~~0.90~~$ & $~~-0.382 \pm 0.039~~$ \\
$~~0.85~~$ & $~~-0.385 \pm 0.045~~$ \\
$~~0.80~~$ & $~~-0.384 \pm 0.058~~$ \\
$~~0.75~~$ & $~~-0.411 \pm 0.072~~$ \\
\hline
\end{tabular}
\end{center}
\caption{Values of the parameter $\alpha_N$ for the normal
   to the three-jet plane with $T < T_C$ at $\sqrt{s}$ = 1 TeV.}
\end{table}

%
%
\section{Conclusions}

Some realistic extensions of the Standard Model of elementary particle physics
suggest the existence of a new  massless neutral gauge boson, denoted as
the paraphoton $\gamma^{\prime}$ in this study. This particle is similar to the
ordinary photon, but the couplings of the $\gamma^{\prime}$ 
are very distinct: interactions with SM fermions are negligible
except those with the top quark. Hence, if the paraphoton is radiated off
the top the signature of $\gamma^{\prime}$ events in the channel
$e^+e^- \to t~\bar{t}~\gamma^{\prime}$
consists of a pair of acoplanar top quark jets with missing transverse energy, ~\met,
carried away by the paraphoton. 
Only the all-hadronic top decay mode was selected to ensure a high signal-to-background
ratio and to avoid complications due to final state neutrinos in leptonic W decays.

Based on a multivariate search strategy
prospects to discover the $\gamma^{\prime}$ at the ILC are studied.
This method was necessary   
to pursue because large $t \bar{t} (\gamma)$ SM background,
small signal event rates and little discrimination power
of variables restricted an effective signal selection by the method of consecutive cuts.
Maximizing the probability of each event to be
a paraphoton candidate, 49\% (76\%) of the signal (S) at 0.5 (1) TeV was selected
and the background (B) strongly suppressed, resulting to a $S/\sqrt{B}$
larger than 150 at $\sqrt{s} = 1$ TeV. Allowing for a 5$\sigma$
$\gamma^{\prime}$ discovery significance,
limits on the paraphoton-top quark 'coupling' $M/\sqrt{c_t}$ were derived.
Assuming that the SM provides the only source of background, $e^+e^-$ collisions at 1 TeV
will bound this parameter to $M/\sqrt{c_t} \lsim 0.61$ TeV, which
seems to be the most stringent limit accessible at the next generation 
colliders since huge background expected at the LHC would prevent an improved number.

For the sake of demonstration and simplicity two angular variables,
the Ellis-Karliner angle and the polar angle of the normal 
to the $t~\bar{t}~\gamma^{\prime}$ plane as a function of a thrust cut-off,
were studied to establish the vector nature of the $\gamma^{\prime}$.
After the cut ~\met $>$ 330 GeV to improve the purity of the signal sample
and $1 - x_1 > 0.05$ 
for the cosine of the  Ellis-Karliner angular distribution, 
with $x_1$ the fractional energy of the fastest parton,
both angular distributions are in accord with the spin-1 assignment
of the paraphoton and inconsistent with e.g. a scalar hypothesis.

\subsubsection*{Acknowledgment} 

The work of E.B. and V.B. is partly supported by the grant
NS.1685.2003.2 of the Russian Ministry of Education and Science.
V.B. also acknowledges  support of grant of the "Dynasty" Foundation.
E.B. and V.B. are grateful to DESY and Fermilab for the kind hospitality.
We thank Bogdan Dobrescu for valuable discussion and reading of the manuscript.



\end{document}